\documentclass[12pt]{article}

\usepackage{graphicx}
\usepackage{a4wide}
\usepackage{mathrsfs}
\usepackage{amsmath}
\usepackage{amssymb}
\usepackage{amsfonts}
\usepackage{dcolumn}

\begin{document}

\title{AdS/CFT Aspects of the Cosmological QCD Phase Transition}
\author{
Cong-Xin Qiu\footnote{E-mail: congxin.qiu@gmail.com; Homepage: http://oxo.lamost.org/}\\
\normalsize{\textit{Department of Astronomy, Nanjing University}}\\
\normalsize{\textit{Nanjing, Jiangsu 210093, P. R. China}}\\
}
\date{~}

\maketitle

\begin{abstract}

Recently, deeper understanding of QCD emerges from the study of the
AdS/CFT correspondence. New results include the properties of
quark-gluon plasma and the confinement/deconfinement phase
transition, which are both very important for the scenario of the
QCD phase transition in the early universe. In this paper, we study
some aspects of how the new results may affect the old calculations
of the cosmological QCD phase transition, which are mainly based on
the studies of perturbative QCD, lattice QCD and the MIT bag model.

\vspace{0.5cm} \noindent \emph{Key words}:
    Confinement/deconfinement phase transition;
    Cosmological phase transition;
    Finite temperature QCD;
    Gauge/string duality
\end{abstract}


\section{Introduction\label{sec:introduction}}

Phase transitions can produce relics, affect the anisotropies of the
universe, or have other observable consequences; hence, it is very
important in astrophysics. A particularly important phase transition
is the QCD confinement/deconfinement phase transition, in which the
deconfined quark-gluon plasma (QGP) phase transits to the confined
hadronic phase. By assuming this phase transition is first order and
that it has nonzero surface tension, it suffers chronologically the
processes of supercooling, reheating, bubble nucleation, and may
produce relics such as quark nuggets. For the up to date reviews of
the cosmological QCD phase transition,
see~\cite{Schwarz:2003du,Boyanovsky:2006bf}.

Recently, deeper understanding of QCD emerges from the study of the
anti-de Sitter/conformal field theory (AdS/CFT)
correspondence~\cite{Aharony:1999ti}. In its prototype version, type
IIB superstring theory on $AdS_5 \times S^5$ is dual to $\mathcal{N}
= 4$ $U(N_\mathrm{c})$ super Yang-Mills (SYM) theory in
$(3+1)$-dimensional spacetime~\cite{Maldacena:1997re}. Generally
speaking, conventional quantum field theories make sense only in the
perturbative regions, where the 't Hooft coupling $\lambda_4 =
g_\mathrm{YM}^2 N_\mathrm{c}$ is small; however, the dual gravity
theory is easy to handle when the supergravity (SUGRa) description
becomes reliable, that is, in the strong coupling region. Hence, we
can use the AdS/CFT correspondence to study field theory in the
region where perturbative approaches are not applicable. It is also
believed that the generalization of the AdS/CFT correspondence can
realize some more realistic systems, such as QCD-like theories or
QCD itself, which have running coupling constants (hence are not
conformal), fundamental matter, and reproduce some phase
transitions. Recent reviews on the connection between string theory
and QCD can be found in~\cite{Peeters:2007ab,Mateos:2007ay}.

New observational results from the relativistic heavy ion collider
(RHIC)~\cite{Gyulassy:2004zy} data tell us that the shear viscosity
of the hot plasma is very small~\cite{Teaney:2003kp}; thus the QGP
at temperature $T \gtrsim T_\mathrm{dec}$ should in fact be strongly
coupled~\cite{Shuryak:2003xe,Shuryak:2004cy}, rather than asymptotic
free as we used to think about it, where $T_\mathrm{dec}$ is the
critical temperature of the confinement/deconfinement phase
transition. Hence, all phenomenological applications of QGP, which
are based on perturbative QCD or the MIT bag
model~\cite{Chodos:1974je,DeGrand:1975cf}, should be reconsidered.
These applications include the neutron stars/quark stars and the
cosmological QCD phase transition. AdS/CFT provides an excellent
tool to study them. Because of its strong interactive nature, it can
explore the properties of QGP and the confinement/deconfinement
phase transition, in both the high temperature and high baryon
number density regions. However, in this paper, we will limit our
focus on the property of high temperature region, which is important
for the cosmological QCD phase transition.

The organization of this paper is as follows. In Sec.~\ref{sec:QGP},
we present the results of QGP and the confinement/deconfinement
phase transition from AdS/CFT. In Sec.~\ref{sec:cosmology}, we study
how these new results affect the conventional scenarios of the
cosmological QCD phase transition, including the nucleation rate,
the supercooling scale and the mean nucleation distance. We
summarize our results in Sec.~\ref{sec:conclusion}. We will set
$\hbar = c = k = 1$ throughout this paper.

\section{The Thermodynamical and Hydrodynamical Quantities of QGP Reconsidered\label{sec:QGP}}

\subsection{Entropy, Free Energy, Energy and Pressure}

\subsubsection{$\mathcal{N} = 4$ SYM theory}

The gauge fields of large-$N_\mathrm{c}$ $\mathcal{N} = 4$ SYM
theory are described as open strings ending on $N_\mathrm{c}$
Dirichlet 3-branes (D3-branes). In the large 't Hooft coupling
limit, $\lambda_4 \gg 1$, the entropy density can be calculated from
the Bekenstein-Hawking entropy~\cite{Bekenstein:1973ur} of
non-extremal D3-branes with Ramond-Ramond charge (RR-charge)
$N_\mathrm{c}$, which is~\cite{Gubser:1996de}
\begin{equation}
    s = \frac{\pi^2}{2} N_\mathrm{c}^2 T^3 \mbox{,}
    \label{eqn:entropy__4SYM}
\end{equation}
where $T$ is identified with the Hawking temperature of the black
brane. The result is only $3/4$ to that of the free gas case $s_0 =
(2\pi^2/3) N_\mathrm{c}^2 T^3$. It was argued that the entropy
density can also be calculated from the action $I$ by $V f = T I = V
\epsilon - T V s$~\cite{Witten:1998zw}, thus $f = - (\pi^2/8)
N_\mathrm{c}^2 T^4$~\cite{Gubser:1998nz}. The sound mode dispersion
relation of hydrodynamical calculations in the strongly coupled
limit gives $c_\mathrm{s}^2 = \partial P / \partial \epsilon = 1/3$
and $P + \epsilon = T s$~\cite{Policastro:2002tn}; hence, both the
energy density $\epsilon$ and the pressure $P$ in the strong
coupling case, should be only $3/4$ to the value of weakly coupled
case, which is consistent with the free energy result from the
action $I$. In fact, all CFTs' have similar equation of states
(EoS's) up to some numerical factors~\cite{Gubser:1999vj}, and what
we presented above is just a trivial example.

For the case with not-so-strong coupling, the leading correction is
calculated from the action $I$, which reads~\cite{Gubser:1998nz}
\begin{equation}
    s = s_0 \left[ \frac{3}{4} + \frac{45}{32} \zeta (3) (2 \lambda_4)^{-3/2} + \ldots \right]
        \mbox{,}
\end{equation}
comparing to the weakly coupled case~\cite{Kim:1999sg,Nieto:1999kc}
\begin{equation}
    s = s_0 \left[ 1 - \frac{3}{2 \pi^2} \lambda_4 + \frac{3 + \sqrt{2}}{\pi^2} \lambda_4^{3/2} + \ldots \right]
        \mbox{.}
\end{equation}
The $3/4$ factor reveals the intrinsic difference between strongly
and weakly coupled system.

\subsubsection{The QCD-like Theories\label{subsubsec:thermo_QCD_like}}

However, CFTs are very different from QCD in many aspects. For
example, (i) their coupling constant $\lambda_4$ does not run, hence
they experience no conventional phase transitions, and (ii) they can
only describes fields in the adjoint (color) but not in fundamental
(flavor) representation of the gauge group. The
confinement/deconfinement phase transition is always understood as a
Hawking-Page phase transition~\cite{Hawking:1982dh} between two
background metrics with different free energy density
$f$~\cite{Witten:1998zw} (except the scenario
of~\cite{Gubser:2008ny,Gubser:2008yx,Gubser:2008sz}). The free
energy density of the system can be calculated from the volume of
spacetime $\int d^D x \sqrt{g}$, and the stable spacetime
configuration has the lowest $f$. Flavors are often added by
$N_\mathrm{f}$ spacetime filling (flavor)
branes~\cite{Karch:2000gx,Karch:2002sh}; however, calculations can
be done only in the probe limit (exact quenched approximation),
$N_\mathrm{f} \ll N_\mathrm{c}$. Many efforts have been spent to
construct a more QCD-like dual theory. As a phenomenological
discussion of their applications to cosmology here in this paper, we
do not want to compare their similarity and dissimilarity in detail;
however, to make our results more concrete, we do not limit our
discussion to some special model. We will reveal the bottom-up way
(the AdS-QCD approaches) including the
hard-wall~\cite{Erlich:2005qh,DaRold:2005zs} and
soft-wall~\cite{Karch:2006pv} models, the top-down way including the
D3-D7
system~\cite{Babington:2003vm,Kirsch:2004km,Ghoroku:2005tf,Apreda:2005yz}
and the D4-D8-$\overline{\mathrm{D8}}$ system (the Sakai-Sugimoto
model)~\cite{Sakai:2004cn,Sakai:2005yt}, and also some other
phenomenological approaches. The comparative theories include the
MIT bag model~\cite{Chodos:1974je,DeGrand:1975cf}, the fuzzy bag
model~\cite{Pisarski:2006hz} and some lattice results. Most gravity
dual theories are limited to the large-$N_\mathrm{c}$ limit;
however, our QCD has $N_\mathrm{c} = 3$, which makes quantitative
applications of the AdS/CFT results difficult. We will try to
compare the disagreement between $N_\mathrm{c} \rightarrow \infty$
and $N_\mathrm{c} = 3$ by some lattice results~\cite{Lucini:2005vg}.
Because of the context of this study, we will always assume that the
chemical potential $\mu = 0$ in this paper, hence the relation
between the free energy density and the pressure is $f = -p$.

Let us first discuss the AdS/QCD approaches. In the hard-wall model,
a cutoff is set in the infrared (IR) region to form a slice of
$AdS_5$, which makes the boundary theory
confining~\cite{Erlich:2005qh,DaRold:2005zs}. The two solutions of
the Einstein equation are a cutoff thermal AdS and a cutoff AdS with
a black hole. For the Ricci flat horizon
case~\cite{Herzog:2006ra,BallonBayona:2007vp}
\begin{equation}
    f_\mathrm{q} - f_\mathrm{h} = \left\{
    \begin{array}{ll}
        (\pi^4 L^3 / 2 \kappa_5^2) T^4 & T < 2^{-1/4} T_\mathrm{dec} \\
        - (\pi^4 L^3 / 2 \kappa_5^2) (T^4 - T_\mathrm{dec}^4) & T >
        2^{-1/4} T_\mathrm{dec}
    \end{array}
    \right. \mbox{,}
    \label{eqn:free_energy__hard-wall_flat}
\end{equation}
where the subscript $\mathrm{h}$ indicates the confining phase,
$\mathrm{q}$ indicates the deconfining phase, $\kappa_5^2 = 8 \pi
G_5$ describes the gravitational coupling scale, and $L$ is the
radius of the AdS space. For the spherical horizon case with
sufficient small IR cutoff $r_0$, we have~\cite{Cai:2007zw}
\begin{equation}
    f_\mathrm{q} - f_\mathrm{h} =
        - \frac{2 \pi^2 \Omega_3}{9 \kappa_5^2} T_\mathrm{dec}^2
        \left( r_+^4 - 2 r_0^4 - \frac{9 r_+^2}{4 \pi^2 T_\mathrm{dec}^2} \right)
    \mbox{,}
    \label{eqn:free_energy__hard-wall_spherical}
\end{equation}
where $\Omega_3 = 2 \pi^2$ and $r_+ = (3 / 8 \pi T_\mathrm{dec})
(\sqrt{9 T^2 / T_\mathrm{dec}^2 - 8} + 3 T/T_\mathrm{dec})$. The
latter case has little physical applications; however, it has
thermodynamical properties similar to the soft-wall case.

In the soft-wall model, the IR cutoff is replaced by a smooth cap
off, which is realized by the dilaton term in the Einstein
action~\cite{Karch:2006pv}. The difference of the free energy
density of the two phases is~\cite{Herzog:2006ra}
\begin{equation}
    f_\mathrm{q} - f_\mathrm{h} =
        \frac{\pi^4 L^3}{\kappa_5^2} T^4 \left[ e^{-x} (x-1) + \frac{1}{2} + x^2 \mathrm{Ei}(-x) \right]
    \mbox{,}
    \label{eqn:free_energy__soft-wall}
\end{equation}
where $x = (T_\mathrm{dec} / 0.491728 \pi T)^2$, and
$\mathrm{Ei}(-x) = -\int_x^\infty e^{-t}/t dt$.

For a ten-dimensional ``AdS/QCD cousin'' model with the metric of a
deformed $AdS_5$ black hole crossing some $5$-dimensional compact
space~\cite{Andreev:2007zv}, the free energy density is
\begin{equation}
\begin{split}
    f_\mathrm{q} - f_\mathrm{h} = & - \frac{\hat{s}}{4} T^4 \left\{
    \left( 1 - \frac{T_\mathrm{dec}^2}{T^2} \right)
    + \left[ - \frac{1}{4} \frac{T_\mathrm{dec}^4}{T^4} \ln \left( \frac{T_\mathrm{dec}^2}{T^2}
    \right) \right. \right. \\
    & \left. \left. - 0.039 \frac{T_\mathrm{dec}^4}{T^4}
    + \sum_{n=3}^{\infty} \frac{(-1)^n}{2^{n-1} (2 - n) n!}
    \left( \frac{T_\mathrm{dec}^2}{T^2} \right)^n \right]
    \right\}
    \mbox{,}
    \label{eqn:free_energy__10dimCousin}
\end{split}
\end{equation}
which is related to a entropy density $s = \hat{s} T^3
\exp{(-T_\mathrm{dec}^2 / 2 T^2)}$. This model may be applicable to
QCD for $1.2 T_\mathrm{dec} < T < 3 T_\mathrm{dec}$. It has a good
asymptotic behavior $\lim_{T\rightarrow\infty} s \propto T^3$ as a
four-dimensional thermal system, because the contributions of the
Kaluza-Klein modes are not taken into account. When $T_\mathrm{dec}
\ll T$, the result coincides the fuzzy bag
model~\cite{Pisarski:2006hz} in pure glue case, which restricts
$B_\mathrm{fuzzy} = f_\mathrm{pert} T_\mathrm{dec}^2 = (\hat{s}/4)
T_\mathrm{dec}^2$, $B_\mathrm{MIT} = 0$ hence $f_\mathrm{q} -
f_\mathrm{h} = - (\hat{s}/4) T^4 (1 - T_\mathrm{dec}^2 / T^2)$.

There are also some other models, like the one defined by some
complex metric in~\cite{Kajantie:2006hv}, the one include a
nontrivial dilaton flow deformation~\cite{Evans:2008tu}, or the MIT
bag model itself. They all have $f_\mathrm{q} - f_\mathrm{h} \propto
T_\mathrm{dec}^4 - T^4$, hence are identical with each other up to
an overall constant. And in fact, for small supercooling, they are
much similar to what in Eq.~(\ref{eqn:free_energy__hard-wall_flat}).

Next, we will discuss the top-down scenarios. In the D3-D7
system~\cite{Babington:2003vm,Kirsch:2004km,Ghoroku:2005tf,Apreda:2005yz},
$N_\mathrm{c}$ coincident D3-branes form an extremal black brane
with near horizon geometry $AdS_5 \times S^5$, while $N_\mathrm{f}$
coincident probe D7-branes fill $AdS_5$ (hence, they also extend
along the radial direction) and wrap some $S^3$ inside $S^5$. When
the D7-branes are separated from the D3-branes in $S^5$, the chiral
symmetry and conformal invariance are broken. When the temperature
is low, the separation is large enough that the brane tension can
avoid the D7-branes falling into the black brane, hence the branes
are ``Minkowski'' embedded outside the horizon. However, when the
temperature is high enough, the gravitational attraction of the
black brane renders the D7-branes a ``black hole''
embedding~\cite{Mateos:2006nu,Mateos:2007vn}. The critical
temperature is $T_\mathrm{fund}$, where the mesons melt. The
multi-valued nature of the free energy density makes the phase
transition first order. Nevertheless, for massive fundamental
quarks, it is not the temperature of the confinement/deconfinement
phase transition, which occurs at some $T_\mathrm{dec} <
T_\mathrm{fund}$. There is as yet a lack of suitable models of
confinement/deconfinement phase transition within D3-D7 system. The
explicit solutions of $f(T)$, $s(T)$ and $c_\mathrm{s}(T)$ are shown
numerically in~\cite{Mateos:2006nu,Mateos:2007vn}. For our purpose,
we will not discuss this ``melting'' transition in detail;
notwithstanding, we take notice of some of its critical parameters
which can be compared to that in the confinement/deconfinement phase
transition. The discontinuity of the entropy density in the phase
transition point is
\begin{equation}
    \Delta s (T = T_\mathrm{fund}) \simeq 0.066 \times \frac{\lambda_4 N_\mathrm{c}
        N_\mathrm{f}}{32} T_\mathrm{fund}^3
        \simeq 0.032~\frac{T_\mathrm{fund}^3}{T^3}~\lim_{T \rightarrow \infty} s_\mathrm{fund}
    \mbox{,}
\end{equation}
which is proportional to $N_\mathrm{c} N_\mathrm{f}$, because only
the contribution of the fundamental matter is taken into account.
The entropy density of massless quarks is $\lim_{T \rightarrow
\infty} s_\mathrm{fund} = \lambda_4 N_\mathrm{c} N_\mathrm{f} T^3 /
16$, and the entropy density attributed to gluons is as what in
Eq.~(\ref{eqn:entropy__4SYM}). The superheating and supercooling
ranges are (by the system itself rather than by impurities or
perturbations)
\begin{equation}
    \Delta_< = 1 - \frac{T_\mathrm{min}}{T_\mathrm{fund}}
    \simeq 0.0019
    ~~\mbox{and}~~
    \Delta_> = \frac{T_\mathrm{max}}{T_\mathrm{fund}} - 1
    \simeq 0.0083\mbox{.}
\end{equation}
The speed of sound also deviates from $1/\sqrt{3}$ nontrivially when
$T$ approaches $T_\mathrm{fund}$. However, unless in the extreme
supercooling case, $c_\mathrm{s}$ would not be vanishing.

Does this ``melting'' transition happens in QCD? This is an
intractable question. Even if we neglect the influences of the
large-$N_\mathrm{c}$ and the probe simplifications, we will still
need QGP remaining strongly coupled at $T_\mathrm{fund}$; because
when it is weakly coupled, the melting of the mesons should be a
crossover. As we will see later, the numerical values of $\Delta s$,
$\Delta_<$ and $\Delta_>$ are all much smaller than the typical
confinement/deconfinement case; besides, we don't really know how to
estimate the surface tension $\sigma_\mathrm{fund}$ of this phase
transition. In addition, melting of different mesons may be
asynchronous in QCD. Of course, if it is indeed a phase transition
in QCD, it can also affect the evolution of our universe.

In the Sakai-Sugimoto model~\cite{Sakai:2004cn,Sakai:2005yt}, when
the temperature is low enough, the $N_\mathrm{c}$ coincident
D4-branes are compactified on a supersymmetry-breaking spacelike
$S^1$ to make the low energy QCD-like theory $(3+1)$-dimensional,
while the $N_\mathrm{f}$ D8-$\overline{\mathrm{D8}}$ pairs (with D8
and $\overline{\mathrm{D8}}$-branes coincide respectively) cross the
$S^1$ circle at some characteristic points. Gauge bosons are
regarded as massless modes of open strings with both ends on
D4-branes, while fundamental fermions correspond to open strings
with one end in some D4-brane and another end in some D8 or
$\overline{\mathrm{D8}}$-brane. However, when the temperature is
high enough~\cite{Aharony:2006da}, to make a lower free energy, the
compactified D4-brane direction is not spacelike but in fact
timelike. This is the confinement/deconfinement phase transition,
because the topological change of the spacetime makes the
expectation value of a temporal Wilson loops change from $\langle
W(C) \rangle = 0$ to $\langle W(C) \rangle \neq 0$. The spontaneous
chiral symmetry breaking is understood as when the $N_\mathrm{f}$
D8-branes and $N_\mathrm{f}$ $\overline{\mathrm {D8}}$-branes merge
at some radial position $u_0$ away from the horizon (where we live),
which happens at some temperature higher or equal to
$T_\mathrm{dec}$. The difference between the free energy densities
of the two phases can be calculated from the DBI action. This phase
transition is first order,
\begin{equation}
    f_\mathrm{q} - f_\mathrm{h} =
        - \frac{40960 \pi^{11}}{729}
        \frac{l_\mathrm{s} (g_\mathrm{s} N_\mathrm{c}) N_\mathrm{c}^2}{T_\mathrm{dec}}
        (T^6 - T_\mathrm{dec}^6) \mbox{.}
        \label{eqn:free_energy__Sakai-Sugimoto}
\end{equation}
For its $AdS_6$ non-critical string ``cousin''
model~\cite{Mazu:2007tp}, $f_\mathrm{q} - f_\mathrm{h} \propto -
N_\mathrm{c}^2 (T^5 - T_\mathrm{dec}^5)$. In the Sakai-Sugimoto
model, one always have the speed of sound $c_\mathrm{s} =
1/\sqrt{5}$~\cite{Benincasa:2006ei}. Because $l_\mathrm{s}
(g_\mathrm{s} N_\mathrm{c}) = g_5^2 N_\mathrm{c} / (2 \pi)^2 = g_4^2
N_\mathrm{c} / (2 \pi)^3 T_\mathrm{dec} = \lambda_4 / (2 \pi)^3
T_\mathrm{dec}$, and $\lambda_4 N_\mathrm{c} / 216 \pi^3 \simeq 7.45
\times 10^{-3}$ from meson spectrum~\cite{Sakai:2005yt}, we see that
the coefficient of Eq.~(\ref{eqn:free_energy__Sakai-Sugimoto}) is
really huge. However, these results are quantitatively far from QCD;
the unwanted Kaluza-Klein modes of the compactified dimensions cause
the theories lacking of the asymptotic UV behavior $f \propto
N_\mathrm{c}^2 T^4$ while $T \rightarrow \infty$. As we will scale
all these theoretical models to QCD by their high temperature
behavior, we will not consider the Sakai-Sugimoto model from now on.

There are also some more phenomenological approaches to the EoS's of
the QCD-like theories. G\"{u}rsoy \emph{et al.} considered a
five-dimensional gravity theory coupled to a dilaton
field~\cite{Gursoy:2007cb,Gursoy:2007er}. The thermodynamics of this
system can be determined uniquely by a positive and monotonic
potential $V(\lambda) = 12 [1 + \lambda + V_1 \lambda^{2Q} \log{}^P
(1+V_2 \lambda^2)]$, where $\phi = \log{\lambda}$ is the dilaton
field~\cite{Gursoy:2008bu,Gursoy:2008za}. The theory is confined
when $Q = 2/3$ and $P > 0$, or $Q > 2/3$. After chosen some specific
potential, the temperature is fixed uniquely by the horizon value of
$\lambda$, and the EoS can be given by some numerical calculations
of the black hole configuration while varying $\lambda(r_H)$. The
aim of this model is still limited to explain the finite temperature
large-$N_\mathrm{c}$ Yang-Mills theory by these authors; however, we
may expect that can tell us something more about QCD.

Gubser \emph{et al.} considered another five-dimensional gravity
theory coupled to a single scalar~\cite{Gubser:2008ny}. Based on a
lot of assumptions, it is shown that the potential of a scalar field
$V(\phi)$ and the EoS of the boundary theory have one-to-one
correspondence. The results may be applicable for regions both
$\gtrsim$ and $\lesssim T_\mathrm{dec}$. Various $V(\phi)$'s
correspond to different EoS's, include crossover, first order, and
second order phase transitions; hence, the authors expect their
model can mimic the EoS of QCD. We choose in this study the
potential $V(\phi) = [-12 \cosh{(\gamma \phi + b \phi^2)}] / L^2$
with $\gamma = \sqrt{7/12}$ and $b = 2$ for the first order case
when making comparison with other models~\footnote{Our EoS's agree
with~\cite{Gubser:2008ny} qualitatively, and can catch all the
expected limits; however, it has some small quantitative divergences
rising from the numerical trickings. These divergences can affect
the critical parameters we choose for the second order phase
transition, but will not affect the main results of our
discussions.}, but keep $\gamma$ as a free parameter in the
discussion of Sec.~\ref{subsubsec:extremely_weak_1st_order}.

The phenomenological model in~\cite{Gursoy:2007cb,Gursoy:2007er}
coupled to the dilaton potential $V(\phi)$ has a more solid
theoretical foundation; however, the calculation of the EoS's is
more complicated than the latter one. As we need to exploit a whole
family of EoS's for our astrophysical purpose (especially in
Sec.~\ref{subsubsec:extremely_weak_1st_order}), we will limit our
discussion to the latter model. It should be noteworthy to review
the astrophysical application of the first model, especially after
some quantitative comparisons between it and the lattice results
that have been done.

\begin{figure}[ht]
\begin{center}
\includegraphics[angle=0,width=9cm]{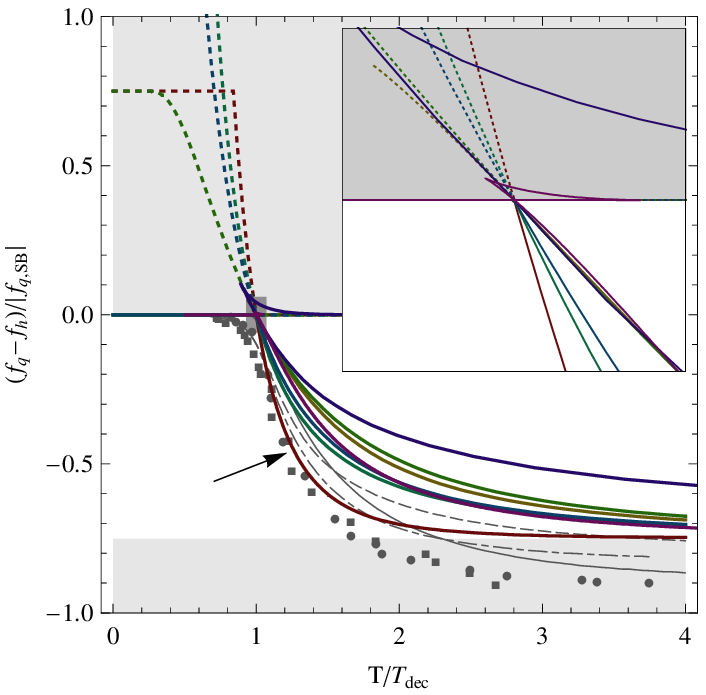}
\end{center}
\caption{The free energy density for various models compares to the
free gas case. For the reason that all gravity dual theories are
strongly coupled, we scale $(\bullet) / (\bullet)_\mathrm{q,~SB}$ to
$3/4$ as $T \rightarrow \infty$, based on the $\mathcal{N} = 4$ SYM
result in Eq.~(\ref{eqn:entropy__4SYM}), where $(\bullet)$ can be
replaced by any thermodynamical quantities, such as entropy, free
energy, energy or pressure. The MIT bag
model~\cite{Chodos:1974je,DeGrand:1975cf} and the fuzzy bag
model~\cite{Pisarski:2006hz} are also scaled to $3/4$ by some
comparison reasons; they can be easily transform back to their
original form if needed. For clarity, we classify and tag our models
by numbers. From the arrow direction marked in this figure, the
thick lines are for the models (1 $\rightarrow$ 4 $\rightarrow$ 5
$\rightarrow$ 7 $\rightarrow$ 2 $\rightarrow$ 3 $\rightarrow$ 6)
respectively. Line(Model)~(1) denotes the hard-wall model with the
Ricci flat horizon calculated in
Eq.~(\ref{eqn:free_energy__hard-wall_flat}), models considered
in~\cite{Kajantie:2006hv,Evans:2008tu}, and the MIT bag model
itself. We neglect their divergent when $T < 2^{-1/4}
T_\mathrm{dec}$. Line~(2) denotes the hard-wall model with the
spherical horizon in
Eq.~(\ref{eqn:free_energy__hard-wall_spherical}). Line~(3) indicate
the soft-wall model case, as Eq.~(\ref{eqn:free_energy__soft-wall})
shows. Line~(4) indicate the ten-dimensional ``AdS/QCD cousin''
model in Eq.~(\ref{eqn:free_energy__10dimCousin}). Line~(5) denotes
the fuzzy bag model result for comparison with Line~(4). Line~(6) is
for the G\"{u}rsoy \emph{et al.} model given
in~\cite{Gursoy:2008bu}. Line~(7) is calculated by the
phenomenological model of~\cite{Gubser:2008ny}, with a scalar
potential $V(\phi) = [-12 \cosh{(\sqrt{7/12} \phi)} + 2 \phi^2] /
L^2$. The thin gray lines are the p4-action
result~\cite{Karsch:2000ps}, in which the solid line indicates the
pure glue case, the dashed line for the $(2+1)$ flavor case, and the
dashed-dotted line for the $3$ flavor case. The points are
calculated by the lattice methods with almost physical quark
masses~\cite{Cheng:2007jq}, where small solid bullets for $N_{\tau}
= 4$ case and solid squares for $N_{\tau} = 6$ case. The small dark
region near the critical temperature is enlarged and shown in the
top-right corner, where the triangle-like shape formed by some line
segments shows clearly the multi-valued nature of Line~(6) and~(7).}
\label{fig:T_vs_F}
\end{figure}

\begin{figure}[ht]
\begin{center}
\includegraphics[angle=0,width=9cm]{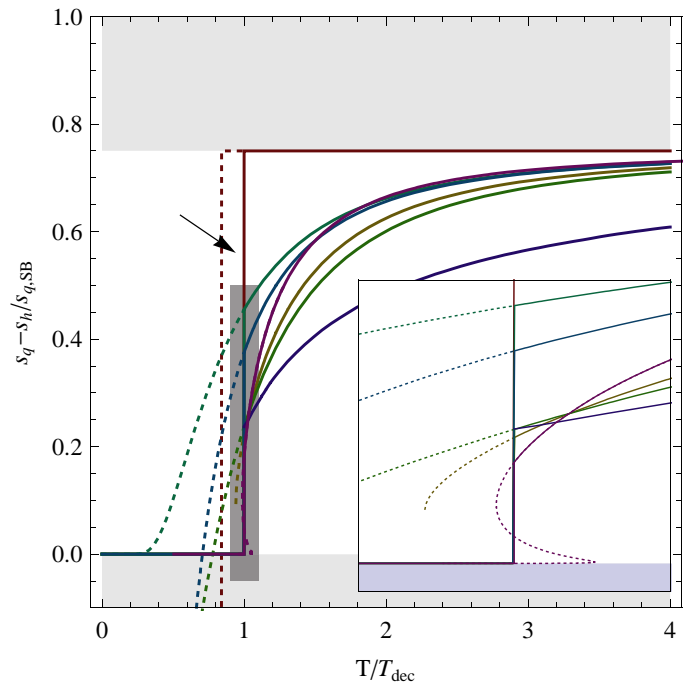}
\end{center}
\caption{The entropy density $s = - d f / d T$ for various models
compares to the free gas case. In fact, in the large-$N_\mathrm{c}$
limit, we have $s_\mathrm{q} \propto N_\mathrm{c}^2$ and
$s_\mathrm{h} \propto N_\mathrm{c}^0$, hence $s_\mathrm{h} = 0$. The
notations are as in Fig.~\ref{fig:T_vs_F}. The thick lines are
models (1 $\rightarrow$ 4 $\rightarrow$ 5 $\rightarrow$ 7
$\rightarrow$ 2 $\rightarrow$ 3 $\rightarrow$ 6) respectively seeing
from the arrow direction.} \label{fig:T_vs_S}
\end{figure}

\begin{figure}[ht]
\begin{center}
\includegraphics[angle=0,width=9cm]{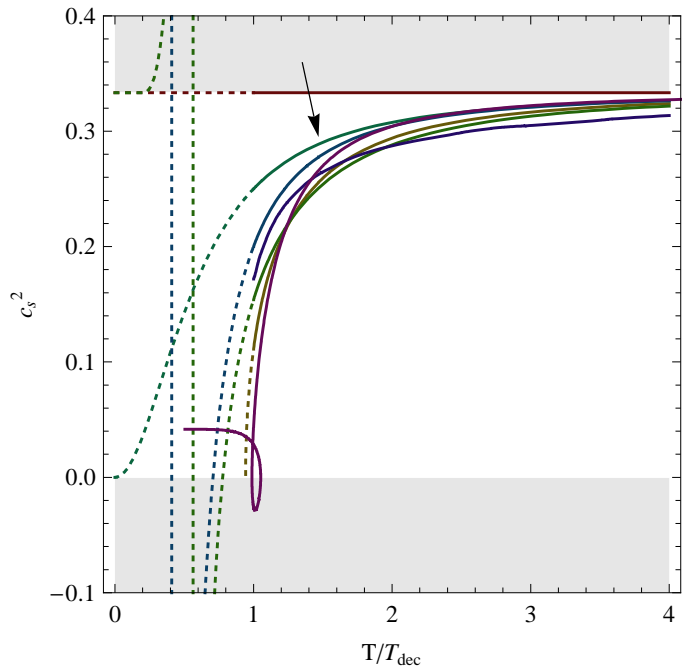}
\end{center}
\caption{The square of sound speed $c_\mathrm{s}^2 = d \log{T} / d
\log{s}$ for various models compares to the free gas case. For the
ideal gas case, or the strongly coupled $\mathcal{N} = 4$ SYM theory
indicated in Eq.~(\ref{eqn:entropy__4SYM}), we have $c_\mathrm{s}^2
= 1/3$. The notations are as in Fig.~\ref{fig:T_vs_F}. The thick
lines are models (1 $\rightarrow$ 4 $\rightarrow$ 5 $\rightarrow$ 7
$\rightarrow$ 6 $\rightarrow$ 2 $\rightarrow$ 3) respectively seeing
from the arrow direction.} \label{fig:T_vs_Cs}
\end{figure}

\begin{figure}[ht]
\begin{center}
\includegraphics[angle=0,width=12cm]{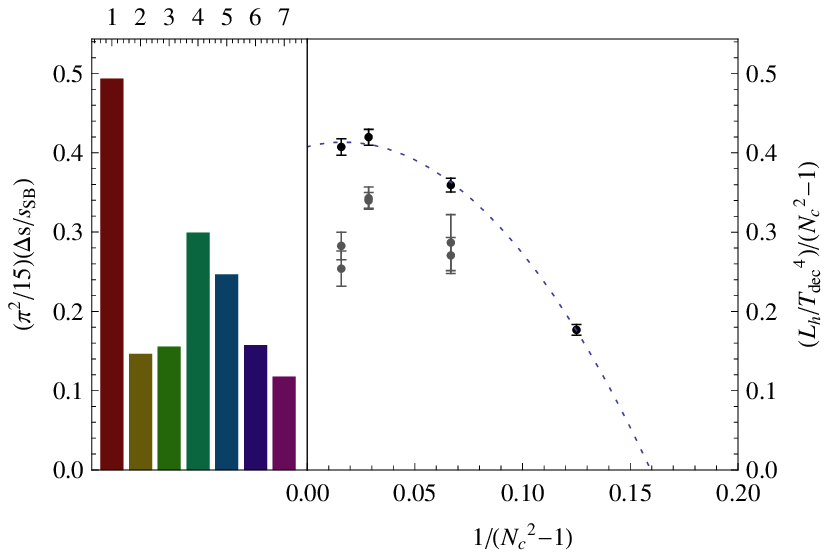}
\end{center}
\caption{The normalization of the latent heat for pure glue fields
of various models. In this case, we have $(\pi^2/15)(\Delta s/
s_\mathrm{SB}) = (L_\mathrm{h} / T_\mathrm{dec}^4) / (N_\mathrm{c}^2
- 1)$. The left part is calculated for various models as tagged in
Fig.~\ref{fig:T_vs_F}. The right part shows that lattice results for
$N_\mathrm{c} = 3, 4, 6$, and $8$~\cite{Lucini:2005vg}, which
suggest that the phase transition is second order for $N_\mathrm{c}
= 2$, weakly first order for $N_\mathrm{c} = 3$, and robustly first
order for $N_\mathrm{c} \geq 4$. The black error bars are for
$L_\mathrm{t} = 5$, and the gray ones for $L_\mathrm{t} = 6$ and
$8$. The fitting line for $L_\mathrm{t} = 5$ case is informal, but
the extend to $N_\mathrm{c} \rightarrow \infty$ case can be guess in
this fitting. Notice that the original MIT bag model has the value
$\pi^2 / 15 \simeq 0.658$ in this figure.} \label{fig:latentHeat}
\end{figure}

The comparisons of the free energy density $f$, the entropy density
$s$ and the square of sound speed $c_\mathrm{s}^2$ for various
models, are shown in Fig.~\ref{fig:T_vs_F},~\ref{fig:T_vs_S}
and~\ref{fig:T_vs_Cs}. We scale all thermodynamical quantities by
$T_\mathrm{dec}$ and $\lim_{T \rightarrow \infty}(\bullet) /
(\bullet)_\mathrm{q,~SB} = 3/4$, where ``SB'' denotes the
Stefan-Boltzmann values of thermal quantities in the corresponding
QGP phase, except the model discussed
in~\cite{Gursoy:2008bu,Gursoy:2008za}. The scaling relation is based
on Eq.~(\ref{eqn:entropy__4SYM}) and the fact that all gravity dual
theories are strongly coupled; we assume that all field considered
are UV conformal, and the coefficient $3/4$ is universally
applicable for them all. The model
in~\cite{Gursoy:2008bu,Gursoy:2008za} is excluded, because it is
indeed weakly coupled in the UV region and asymptotically
Stefan-Boltzmann. The rescaling is of course reasonable for the
entropy $s$ in Fig.~\ref{fig:T_vs_S}, because in the
large-$N_\mathrm{c}$ theories, $s_\mathrm{q} \propto N_\mathrm{c}^2$
and $s_\mathrm{h} \propto N_\mathrm{c}^0$; hence the latter one can
always be neglected. For the free energy $f$, things are a little
more subtle, since the UV cutoff introduced by the
computation~\cite{Witten:1998zw} do not ensure $f_\mathrm{h} \propto
N_\mathrm{c}^0$. We make the statement as an assumption by using
some appropriate counter-terms. Although some models
(e.g.,~\cite{Andreev:2007zv} and~\cite{Gubser:2008ny}) aim directly
at QCD itself, we assume the superheating contributions of
$f_\mathrm{h}$ and $s_\mathrm{h}$ in their models can also be
neglected  comparing to QGP while $T \rightarrow \infty$. The
numbers for the classified models are tagged in
Fig.~\ref{fig:T_vs_F}. In Tab.~\ref{tab:superheating_supercooling},
we list the maximal superheating and supercooling scale $\Delta_>$
and $\Delta_<$ for various confinement/deconfinement models, and
also the ``melting'' transition
in~\cite{Mateos:2006nu,Mateos:2007vn}. The existence of $\Delta_>$
and $\Delta_<$ indicates a completely different phase transition
process comparing to the old one; for the range of superheating or
supercooling is no longer caused by impurities or perturbations, but
caused by the theoretical system itself. It can be seen that the
``melting'' values are much smaller than the
confinement/deconfinement case. In Fig.~\ref{fig:latentHeat}, we
compare the latent heat $L_\mathrm{h}$ from the theoretical models
list above, and from the lattice calculations for various
$N_\mathrm{c} \geq 3$.

\begin{table}[ht]
\caption{Comparison of superheating scale $\Delta_> =
T_\mathrm{max}/T_\mathrm{dec} - 1$ and supercooling scale $\Delta_<
= 1 - T_\mathrm{min}/T_\mathrm{dec}$ for various models. The models
are numbered as in Fig.~\ref{fig:T_vs_F}. The $\Delta_>$ and
$\Delta_<$ in the ``melting''
transition~\cite{Mateos:2006nu,Mateos:2007vn} are simply replaced
$T_\mathrm{dec} $ by $T_\mathrm{fund}$.}
\label{tab:superheating_supercooling}
\begin{center}
\begin{tabular}{l r @{.} l r @{.} l} \hline \hline
Model No. & \multicolumn{2}{l}{$\Delta_>$} & \multicolumn{2}{l}{$\Delta_<$} \\ \hline
1 & \multicolumn{2}{l}{$\infty$} & 0&159~($\infty$) \\
2 & \multicolumn{2}{l}{$\infty$} & 0&057 \\
3 & \multicolumn{2}{l}{$\infty$} & 1& \\
4 & \multicolumn{2}{l}{$\infty$} & 1& \\
5 & \multicolumn{2}{l}{$\infty$} & 1& \\
6 & \multicolumn{2}{l}{$\infty$} & 0&111 \\
7 & 0&046 & 0&011 \\
``melting''($T_\mathrm{fund}$) & 0&0083 & 0&0019 \\ \hline \hline
\end{tabular}
\end{center}
\end{table}

Heuristically, we see that for temperature $2 T_\mathrm{dec}
\lesssim T \lesssim 4 T_\mathrm{dec}$, models~(2), (3) and (4), (5),
(7) look similar to each other; and for temperature $T \simeq
T_\mathrm{dec,+}$, models~(2), (3), (6), (7) look similar.
Models~(2), (3), (6), and (7) have the latent heat a little too
small compared to the lattice result of the large-$N_\mathrm{c}$
theories, and the latent heat of model~(1) and the original MIT bag
model seem too large. The discussions of models~(6) and (7) may be a
little more unreasonable, because the free parameters in the dilaton
potential are chosen arbitrarily. The divergence of these models
rise because they are all quantitatively far from
(large-$N_\mathrm{c}$) QCD. To avoid the unnecessarily complicated
details of these models in our discussions in
Sec.~\ref{sec:cosmology}, it is worthwhile to ask what kind of
feature they have in common. We argue that (i) the EoS of real QCD
should be softer than the bag model, and (ii) there exist some
intrinsic maximum supercooling scale to be achieved, in contrary to
the old belief that the range of supercooling is caused by
impurities or perturbations.

It is interesting to argue in what conditions the bag-model-like
theories can still be applicable. The renormalization-group-improved
perturbation expansion method tells us that, when the strong
coupling constant $\alpha_\mathrm{s}$ increases, the bag constant
$B_\mathrm{MIT}$ decreases~\cite{Farhi:1984qu}. Although this result
is only suitable for the perturbative and zero temperature regions,
it suggests us to treat the bag model carefully. However, there are
indeed a lot of gravity dual theories whose boundary field theories
have bag-model-like
thermodynamics~\cite{Herzog:2006ra,BallonBayona:2007vp,Kajantie:2006hv,Evans:2008tu},
which do well for explaining meson spectrum or other physical
applications.

\subsection{Shear Viscosity and Bulk Viscosity}

\subsubsection{$\mathcal{N} = 4$ SYM theory}

The shear viscosity of large-$N_\mathrm{c}$ $\mathcal{N} = 4$ SYM
theory in the large 't Hooft coupling limit, can be calculated via
the Kubo relations. The result
is~\cite{Policastro:2001yc,Policastro:2002se}
\begin{equation}
    \frac{\eta}{s} = \frac{1}{4 \pi} \mbox{.}
\end{equation}
And for bulk viscosity, conformal property requires $\zeta = 0$. It
was argued that this value is always available for theories with
holographically dual supergravity
descriptions~\cite{Kovtun:2003wp,Buchel:2003tz}. For the case of
large but finite 't Hooft coupling $\lambda_4$, we
have~\cite{Buchel:2004di,Benincasa:2005qc} $\zeta =
\mathcal{O}(\lambda_4^{-3})$ and
\begin{equation}
    \frac{\eta}{s} = \frac{1}{4} \left[ 1 + \frac{135}{8} \zeta (3) (2 \lambda_4)^{-3/2} + \ldots \right]
        \mbox{,}
\end{equation}
which can be compared with the weakly coupled
case~\cite{Huot:2006ys}
\begin{equation}
    \frac{\eta}{s} \simeq \frac{6.174}{\lambda_4^2 \ln{(2.36/\sqrt{\lambda_4})}}
        \mbox{.}
\end{equation}

\subsubsection{The QCD-like Theories}

As for the thermodynamical case, people follow the top-down and
bottom-up routes to discuss the hydrodynamical quantities of
QCD-like theories. However, there is a lack of lattice results to be
compared with, because lattice QCD is incapable for real-time
behaviors.

To break the conformal behavior of AdS/CFT, one easy way is to
consider Dp-branes. The result is~\cite{Mas:2007ng} $c_\mathrm{s}^2
= (5-p)/(9-p)$,
\begin{equation}
    \frac{\eta}{s} = \frac{1}{4\pi} \mbox{~~and~~}
    \frac{\zeta}{\eta} = \frac{2 (3-p)}{p (9-p)} \mbox{.}
\end{equation}
For the case of compactified Dp-brans, the relations for
$c_\mathrm{s}$ and $\eta/s$ are the same as before, but the relation
for $\zeta/\eta$ has to be modified to
\begin{equation}
    \frac{\zeta}{\eta} = \frac{8d - 2(9-p)(d-1)}{d (9-p)}
        = 2 \left( \frac{1}{d} - c_\mathrm{s}^2 \right) \mbox{,}
        \label{eqn:zeta_vs_eta}
\end{equation}
which is consistent with the Sakai-Sugimoto model's result
$\zeta/\eta = 4/15$~\cite{Benincasa:2006ei} for $p = 4$ and $d = 3$.

To take into account the contributions of fundamental matter, one
can consider the D3-D7 system. The result is~\cite{Mateos:2006yd}
\begin{equation}
    \eta = \frac{\pi}{8} N_\mathrm{c}^2 T^3
        \left[ 1 + \frac{\lambda_4}{8 \pi^2} \frac{N_\mathrm{f}}{N_\mathrm{c}}
        h\left( \frac{\lambda_4 T}{M_\mathrm{q}} \right) + \ldots \right]
        \mbox{,}
\end{equation}
where $M_\mathrm{q}$ is the quark mass, $h(x)$ is some smooth
function connects $h(0) = 0$ and $h(\infty) = 1$ by a crossover
around $x \sim 1$, with the entropy density $s = (\pi^2/2)
N_\mathrm{c}^2 T^3 + s_\mathrm{fund}$ already been discussed in
Sec.~\ref{subsubsec:thermo_QCD_like}. Similar calculations for the
Dp-Dq-$\overline{\mathrm{Dq}}$ system including the Sakai-Sugimoto
model can also be done.

For the models of five-dimensional gravity coupled to some dilaton
fields, the bulk viscosity can be calculated directly by the Kubo
formula~\cite{Gubser:2008yx,Gubser:2008sz}. $\zeta$ can be estimated
by the numerical solution of the metric.

Based on the discussions above and also some other evidences, people
conjecture that there may be some universal bounds of shear
viscosity $\eta/s \geq 1/4\pi$ (or $\hbar/4\pi k_\mathrm{B}$ when
getting back the units; also called the Kovtun-Son-Starinets (KSS)
bound) for \emph{all physical systems in
Nature}~\cite{Kovtun:2003wp,Buchel:2003tz,Kovtun:2004de}, and of
bulk viscosity $\zeta/\eta \geq 2 (1/p - c_\mathrm{s}^2)$ for
theories with holographically dual supergravity
descriptions~\cite{Buchel:2007mf}. The universality of these bounds
suggests that we can use them as critical parameters for the
properties of QGP; however, different opinions of them exist in
literatures. Clues from the generalization of the second law of
thermodynamics (GSL) suggests some origin of the KSS bound from very
basic physical principle~\cite{Fouxon:2008pz}; nevertheless, various
theoretical models have being constructed which violate the bound,
both from quantum field
theory~\cite{Cohen:2007qr,Son:2007xw,Cohen:2008zz,Cherman:2007fj}
and from AdS/CFT itself~\cite{Kats:2007mq,Brigante:2007nu}.
Fortunately, the latter violation only loosens the bound a little,
to $\eta/s \geq (16/25) (1/4\pi)$, for the constraint of
causality~\cite{Brigante:2007nu,Brigante:2008gz}. In addition, using
the model constructed in~\cite{Gubser:2008yx,Gubser:2008sz} to
calculate the bulk viscosity of the potential $V(\phi) = [-12
\cosh{(\gamma \phi + b \phi^2)}] / L^2$, can sometimes violate the
bound given in~\cite{Buchel:2007mf}.

For concreteness, we go back to the case of QGP itself. Let us first
discuss the shear viscosity $\eta$. Although some theoretical
arguments suggest us that $\eta/s$ should be much larger (maybe by a
constant of $\sim 7$) than $1/4\pi$ in the strong 't Hooft coupling
limit, because it is much larger than the $\mathcal{N} = 4$ SYM
theory case in the weak coupling limit~\cite{Huot:2006ys}, RHIC
results tell us that the $\eta/s$ of QGP nearly
saturates~\cite{Teaney:2003kp,Gavin:2006xd,Majumder:2007zh}, or
maybe even violates~\cite{Majumder:2007zh} the KSS bound.

There are few discussions about the dependence of parameter $\eta$
on the temperature $T$. It has been done in the hard-wall and the
``AdS/QCD cousin'' models~\cite{Kapusta:2008ng}; nevertheless, they
both always have $\eta/s < 1/4\pi$, which violate the KSS bound.
Na\"{i}vely, one can estimate it by some phenomenological relation
\begin{equation}
    \eta \sim \epsilon l c_\mathrm{s} \mbox{,}
\end{equation}
where $l$ is the correlation length; however, it is very hard to
make quantitative computations by this formula. Some interpolation
between strong and weak coupling regions may be also
possible~\cite{Hirano:2005wx}, as the perturbative QCD result of
$\eta$ in the weak coupling region is rather
credible~\cite{Arnold:2000dr}.

For the case of the bulk viscosity $\zeta$, lattice results of
gluodynamics show that it rises sharply when $T \rightarrow
(T_\mathrm{dec})_+$~\cite{Kharzeev:2007wb,Meyer:2007dy,Karsch:2007jc},
which are qualitatively consistent with the fact that $c_\mathrm{s}$
drops there. Although $\zeta$ cannot be calculated in the
supercooling region $T < T_\mathrm{dec}$ within the lattice
framework, we assume from AdS/CFT that it varies smoothly while
cross the phase transition point.

\subsection{Surface Tension\label{subsec:surface_tension}}

Very few works exist addressing the surface tension of the
confinement/deconfinement phase transition from the AdS/CFT
viewpoint. For this purpose, two separate metrics with different
topologies, both have $(3+1)$-dimensional translational invariance
within ``our world'' (as assumed by all the models in
Sec.~\ref{subsubsec:thermo_QCD_like}), are not suitable; as we need
nontrivial metric change along the direction of ``our world''. Some
relative discussions can be found in~\cite{Aharony:2005bm}.
Deconfined regions map to some pancake-like black hole solutions,
whose interior resembles black brane; however, they have
domain-wall-like boundary to smoothly connect with the confined
gravity solution. Hence, the hadronization of the plasma balls can
be understood as the Hawking radiation of the dual black holes.
Although this work aims particularly at the large-$N_\mathrm{c}$
gauge theories, some other authors believe that dual black holes are
in fact produced inside of
RHIC~\cite{Nastase:2005rp,Shuryak:2005ia}.

The concrete calculation is based on some finite temperature
Scherk-Schwarz compactificational metrics, which have covering space
asymptotically $AdS_{d+2}$ near the boundary. Both the time
direction $\tau$ and a spacelike direction $\theta$ are compactified
to some circles $S^1$; however, the $\theta$ circle shrinks to zero
at some finite $u = u_0$ in the confined phase, rather than the
$\tau$ circle shrinks to zero in the deconfined phase. The metric of
the domain-wall-like boundary can be solved numerically, and the
surface tension can be estimated by it. The surface tension $\sigma$
is rounded to numbers $2.0~\epsilon_\mathrm{q}(T_\mathrm{dec}) /
T_\mathrm{dec}$ for $d = 3$ (a hence $(2+1)$-dimensional gauge
theory) and $1.7~\epsilon_\mathrm{q}(T_\mathrm{dec}) /
T_\mathrm{dec}$ for $d = 4$ (a hence $(3+1)$-dimensional gauge
theory). $\sigma \propto \epsilon_\mathrm{q} \propto N_\mathrm{c}^2$
is a natural result of the scaling of the classical gravity action.
The aftermath of this fact is discussed in
Sec.\ref{subsubsec:large-N_constraint}.

However, there are some relevant discussions of the surface tension
$\sigma$, based on both lattice gauge theory and the MIT bag model.
The lattice results of $\sigma$ for the pure gluon $SU(3)$ gauge
theory are around
$0.02~T_\mathrm{dec}^3$~\cite{Iwasaki:1993qu,Beinlich:1996xg,Lucini:2005vg}.
In the MIT bag model, the contribution of $\sigma$ is divided to an
intrinsic and a dynamical surface tension~\cite{Farhi:1984qu}. The
intrinsic surface tension $\sigma_I$ is suggested to be very small;
however, we do not know how to calculate it in this framework. The
dynamical surface tension $\sigma_D$ raises from the modification of
the fermion density in the phase transition surface; hence, it
depends sensitively on the strange quark mass. Detailed calculation
shows that $\sigma_D$ is at most
$(60~\mathrm{MeV})^3$~\cite{Berger:1986ps}. Notice that the bag
model results are only valid for the zero temperature case, and the
lattice results do not consider fundamental quarks (which is
supposed to be crucial in the bag model discussions). However, these
results may suggest that $\sigma$ is not very large.

\section{The Cosmological QCD Phase Transition Reconsidered\label{sec:cosmology}}

If the QCD confinement/deconfinement phase transition is first
order, just as what the application of a Hawking-Page phase
transition indicates, our universe underwent that transition when it
was about $10^{-5}~\mathrm{s}$ old. Generically, if the surface
tension of the transition interface is nonzero, the universe should
be supercooled for some scale before nucleation indeed
happens~\cite{DeGrand:1984uq,Kajantie:1992uk}. After the
supercooling stage, some hadronic bubbles are created; they may then
expand rapidly as both the
detonation~\cite{Steinhardt:1981ct,Gyulassy:1983rq,Ignatius:1993qn}
and
deflagration~\cite{Gyulassy:1983rq,KurkiSuonio:1984ba,Ignatius:1993qn}
waves. For the deflagration wave case, the latent heat released by
the phase transition, reheats our universe back to $T_\mathrm{dec}$.
After that, the phase transition goes along synchronously while the
universe expands, and converts the denser QGP matter to the
less-dense hadronic matter mildly. The mean distance between the
hadronic bubbles, is calculated
in~\cite{Hogan:1984hx,Fuller:1987ue,Meyer:1991zm,Ignatius:1994fr,Christiansen:1995ic}.
After about half of the QGP matter has been converted, the hadronic
bubbles are replaced by the QGP bubbles. As the phase transition
goes on, the QGP bubbles disappear more and more
rapidly~\cite{Applegate:1985qt}. Baryons may be concentrated in the
QGP bubble, and relics such as quark nuggets may be
produced~\cite{Witten:1984rs}. Some panoramic description of this
phase transition can be found in~\cite{Kajantie:1986hq}, and some up
to date review articles are
in~\cite{Schwarz:2003du,Boyanovsky:2006bf}.

The process we described above is called \emph{homogeneous
nucleation}. We will not consider other possibilities such as
heterogeneous nucleation~\cite{Fuller:1987ue,Christiansen:1995ic} or
inhomogeneous nucleation~\cite{Ignatius:2000cz} in this paper,
because they are less sensitive to the intrinsic properties of QCD
(hence, less sensitive to the AdS/CFT results) than the homogeneous
case. In addition, we will not consider the late stage issues of
this phase transition, such as the stability of quark nuggets,
because the zero chemical potential assumption is no longer suitable
there. We leave the relative discussions in the follow-up studies,
by which the results from the finite chemical potential AdS/CFT
correspondence can be used directly.

\subsection{The Nucleation Rate}

The nucleation rate of the hadronic phase out of the QGP phase can
be calculated as in~\cite{Csernai:1992tj}
\begin{equation}
    \Gamma = \frac{\kappa}{2 \pi} \Omega_0 e^{- \Delta F(R_{*}) / T} \mbox{,}
    \label{eqn:nucleation_rate}
\end{equation}
where
\begin{equation}
    \kappa = \frac{4 \sigma (\zeta_\mathrm{q}
        + 4 \eta_\mathrm{q} / 3)}{T^2 (s_\mathrm{q} - s_\mathrm{h})^2 R_{*}^3}
\end{equation}
is the dynamical prefactor to describe the dissipation effect,
\begin{equation}
    \Omega_0 = \frac{2}{3 \sqrt{3}} \left(\frac{\sigma}{T}\right)^{3/2}
        \left(\frac{R_{*}}{\xi_\mathrm{q}}\right)^4
\end{equation}
is the statistical prefactor, and
\begin{equation}
    \Delta F(R_{*}) = \frac{16 \pi}{3} \frac{\sigma^3}{(f_\mathrm{q} - f_\mathrm{h})^2}
\end{equation}
is the additional free energy of a hadronic bubble of the critical
size $R_{*} = 2 \sigma / (f_\mathrm{q} - f_\mathrm{h})$ within the
QGP phase, $\xi_\mathrm{q}$ is the correlation length in the QGP
phase. For the case of zero chemical potential, we have
$f_\mathrm{q} - f_\mathrm{h} = P_\mathrm{h} - P_\mathrm{q}$ and the
enthalpy density $\omega = s T$.

\begin{figure}[ht]
\begin{center}
\includegraphics[angle=0,width=9cm]{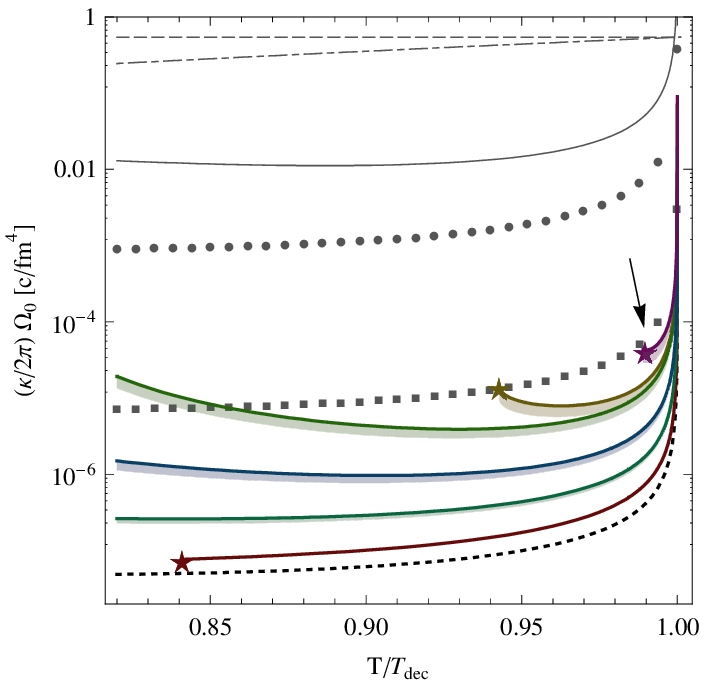}
\end{center}
\caption{The prefactor $(\kappa/2\pi) \Omega_0$ in the nucleation
rate formula. The thin gray dashed and dashed-dotted lines on the
top are for dimensional values $T_c^4$ and $T^4$ respectively. The
thin gray solid line using the same parameters as
in~\cite{Csernai:1992tj}, is shown for comparison reasons. Its value
seems much larger than all other cases, \emph{mainly} because it
uses a rather large $\sigma = 50~\mathrm{MeV}/\mathrm{fm^2}$
(although other parameters also affect the curve); however, we
choose a rather small value of $\sigma = 0.02~T_\mathrm{dec}^3
\simeq 3.64~\mathrm{MeV}/\mathrm{fm^2}$ for $T_\mathrm{dec} =
192~\mathrm{MeV}$~\cite{Cheng:2006qk} in all other estimations. The
gray solid bullet and square lines are for the pure gluon $SU(3)$
lattice result $L_\mathrm{h} = 1.4~T_\mathrm{dec}^4$ and $\sigma =
0.02~T_\mathrm{dec}^3$~\cite{Lucini:2005vg}. Nevertheless, when
calculating the effectively massless degrees of freedom, we also
count the fermionic contributions. The difference is the former case
uses the perturbative result $\eta_\mathrm{q} \simeq 1.12~T^3 /
\alpha_\mathrm{s}^2 \log{(1/\alpha_\mathrm{s})}$ and
$\alpha_\mathrm{s} \sim 0.23$, but the latter case uses the AdS/CFT
result $\eta_\mathrm{q} = s_\mathrm{q}/4\pi$. The thick color lines
are for models discussed above. Seeing from the arrow direction,
they are models (7 $\rightarrow$ 2 $\rightarrow$ 3 $\rightarrow$ 5
$\rightarrow$ 4 $\rightarrow$ 1) respectively. The process of
scaling those large-$N_\mathrm{c}$ theories to real QCD, and the
rationality of that scaling, are discussed in the main text. The
shear viscosity of models (1) and (4) are evaluated
by~\cite{Kapusta:2008ng}; while for all other cases, we choose
$\eta_\mathrm{q} = s_\mathrm{q}/4\pi$. The bulk viscosities are
chosen by the relation $\zeta_\mathrm{q}/\eta_\mathrm{q} = 2 (1/3 -
c_\mathrm{s}^2)$ of Eq.~(\ref{eqn:zeta_vs_eta}), and the shadow
regions show the differences between them and the $\zeta_\mathrm{q}
= 0$ cases. $\zeta_\mathrm{q}$ of model (7) can be calculated from
more sophistical numerical results given
by~\cite{Gubser:2008yx,Gubser:2008sz} if needed; however, we deal
with it similarly with others for simplification. The black dotted
line near the bottom is for the original MIT bag model with
$\eta_\mathrm{q} = (s_\mathrm{q} - s_\mathrm{h})/4\pi$. We choose
the correlation length $\xi_\mathrm{q} =
0.48(T_\mathrm{dec}/T)~\mathrm{fm}$~\cite{Kaczmarek:2004gv} from
lattice result for all our estimations, except the thin gray solid
comparison line; in the gravity side, a lower limit of
$\xi_\mathrm{q}$ is given by~\cite{Fouxon:2008pz}.}
\label{fig:nucleationRatePrefactor_vs_supercooling}
\end{figure}

The prefactor $(\kappa/2\pi) \Omega_0$ in the nucleation rate
formula for various models, is shown in
Fig.~\ref{fig:nucleationRatePrefactor_vs_supercooling}. The most
important step is how to map the various thermodynamical quantities
of large-$N_\mathrm{c}$ theories from AdS/CFT models to real QCD.
Our strategy is linearly map $(\bullet)_\mathrm{q,~SB}$ to the
corresponding quantities of the $g_\mathrm{q} = 37 + 14.25$ ideal
gas model, and map the $f_\mathrm{q} = f_\mathrm{h}$ and
$s_\mathrm{q} = s_\mathrm{h}$ horizontal lines in
Fig.~\ref{fig:T_vs_F} and~\ref{fig:T_vs_S} to the $g_\mathrm{h} = 3
+ 14.25$ ideal gas model, where $g_\mathrm{q}$ and $g_\mathrm{h}$
are the degrees of freedom of the real world at $T = T_\mathrm{dec}
\simeq 192~\mathrm{MeV}$~\cite{Cheng:2006qk} before and after the
confinement/deconfinement phase transition. The coefficient $14.25$,
contributed by the leptons and photons, is almost irrelevant to our
follow up discussions, beside the ones using the Friedmann equations
to describe the expanding universe; hence, we will not discuss its
rationality. However, the contribution $3$ from the pions, actually
needs to be studied more carefully. Pionic freedom is caused by the
fundamental quarks, while $21$ of $37$ in $g_\mathrm{q}$ is caused
by the fundamental quarks as well. As nearly all our models of EoS's
are dominated by gluodynamics, and the contribution to the latent
heat $L_\mathrm{h}$ or the surface tension $\sigma$ by gluons and
quarks cannot be discussed separately, this manipulation is in fact
untenable. However, the quenched lattice method faces the same
problem. Nevertheless, we take the whole EoS's to describe the
thermal quantities in different temperatures, rather than some
characteristic parameters like $L_\mathrm{h}$ or $\sigma$. For some
models with free parameters like
in~\cite{Gubser:2008ny,Gubser:2008yx,Gubser:2008sz} (which we will
discuss especially in
Sec.~\ref{subsubsec:extremely_weak_1st_order}), we may expect that
suitable choice of parameters can absorb the contribution of
fundamental quarks. Hence, we expect the calculations below can
still reveal some aspects of real QCD.

As seen from Fig.~\ref{fig:nucleationRatePrefactor_vs_supercooling},
the strongly coupled nature of QGP can lower the prefactor
$(\kappa/2\pi) \Omega_0$ a lot, mainly by the reason that it has a
relatively smaller shear viscosity $\eta_\mathrm{q} =
s_\mathrm{q}/4\pi$. It is artificial that the lattice results seems
much larger than what is in all of our models tagged by numbers. The
reason is that, the value $L_\mathrm{h} = 1.4~T_\mathrm{dec}^4$ is
calculated by gluodynamics, but it has been shared na\"{i}vely to
both gauge and fundamental particles by our simple mapping. As the
lattice results indicate, the latent heat of QCD with physical
quarks may be smaller than pure gauge case, the prefactor may be
enhanced. The increasing of $(\kappa/2\pi) \Omega_0$ for some
not-very-small supercooling for our models is very interesting.
Beside the reason we erase all the reductions for small
supercooling, the main reason is when the EoS is not bag-model-like,
the latent heat is not as large as in $T_\mathrm{dec}$ while the
supercooling is large. This can be seen roughly from
Fig.~\ref{fig:T_vs_S} and the relation $L_\mathrm{h} = (4/3) T
(s_\mathrm{q} - s_\mathrm{h})$.

\subsection{The Supercooling Scale and the Mean Nucleation Distance}

To estimate the supercooling scale quantitatively, we have some
separate criteria. If the supercooling is required to complete the
phase transition, we need at least one nucleating bubble per Hubble
volume; that is, $\Gamma > 1/d_\mathrm{H}^3 \Delta t$ for the Hubble
radius $d_\mathrm{H} = a/\dot{a} = \sqrt{45/4\pi^3} M_\mathrm{pl}
\cdot g_\mathrm{q}^{-1/2} T^{-2}$ and the nucleating duration
$\Delta t$. We may relax $\Delta t$ to the Hubble time
$d_\mathrm{H}/2$, because the resulting supercooling scale is in
fact insensitive to this parameter. Hence, the supercooling scale
can be roughly estimated by $\Gamma \simeq 1 / d_\mathrm{H}^4$.

To estimate the supercooling scale more accurately, let us consider
the deflagration bubble scenario. The applicable parameter space of
this scenario is discussed in~\cite{Ignatius:1993qn}. Assuming that
a hadronic bubble created in the supercooling QGP phase expands
deflagratingly, a shock wave with velocity $v_\mathrm{sh} \gtrsim
c_\mathrm{s}$ preheats the QGP matter to stop the new nucleating
processes there, and a deflagration wave with relatively slow
velocity $v_\mathrm{def}$ burns the QGP matter to hadronic matter
behind it~\cite{Gyulassy:1983rq,KurkiSuonio:1984ba,Ignatius:1993qn}.
The velocities $v_\mathrm{sh}$ and $v_\mathrm{def}$ are calculated
accurately in~\cite{Ignatius:1993qn}. The weakly and
electromagnetically interacting particles can affect these
velocities~\cite{Miller:1989jr,Miller:1989gj}; however, deflagration
happens only during the early stages for the small supercooling
case, when their influences are negligible. When most of the space
has been swept by the shock wave, the supercooling process ceases.
The fraction of space which has already been swept by the shock wave
is calculated foremost in~\cite{Guth:1979bh,Guth:1981uk}. For our
purpose, we can neglect the expanding of the universe in the
supercooling timescale. Hence, the criterion of the supercooling
scale $T_\mathrm{f}$ is roughly~\cite{Kajantie:1992uk}
\begin{equation}
    \frac{4\pi}{3} \int_{t_\mathrm{dec}}^{t_\mathrm{f}}
        \Gamma v_\mathrm{sh}^3 (t_\mathrm{f} - t)^3 dt \simeq 1 \mbox{,}
\end{equation}
where $t_\mathrm{dec}$($t_\mathrm{f}$) is the age of the universe at
temperature $T_\mathrm{dec}$($T_\mathrm{f}$). This integral equation
can be solved approximately by
\begin{equation}
    \left.
    \left[
        -\frac{\sqrt{24 \pi G} \cdot \epsilon_\mathrm{q}^{1/2} (p_\mathrm{q} + \epsilon_\mathrm{q})}
        {d \epsilon_\mathrm{q} / d T}
        \frac{d(\Delta F/T)}{dT}
    \right]^4
    \simeq
    8 \pi \left(\frac{\kappa}{2\pi} \Omega_0\right) v_\mathrm{sh}^3 e^{-\Delta F/T}
        ~\right|_{T_\mathrm{f}}
        \mbox{,}
        \label{eqn:supercooling_scale}
\end{equation}
in which we deal with the Friedmann equations without any assumption
about the EoS of the QGP phase.

\begin{figure}[ht]
\begin{center}
\includegraphics[angle=0,width=9cm]{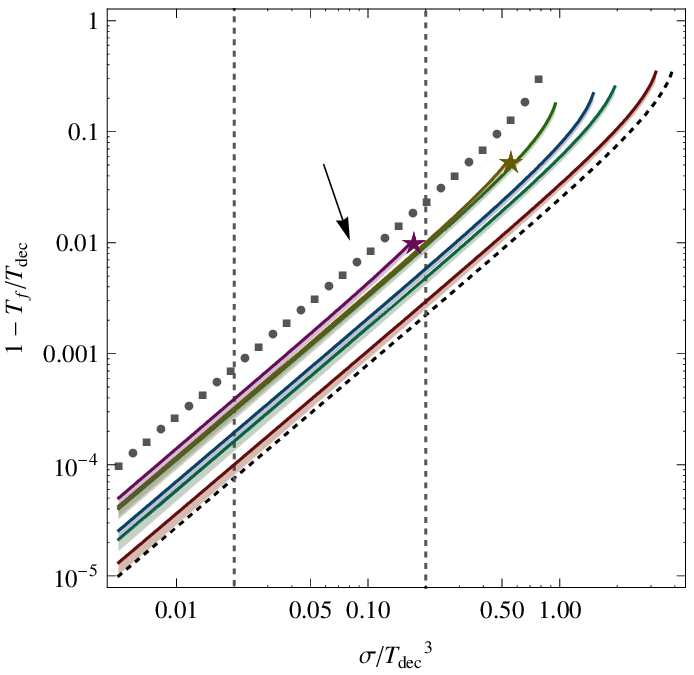}
\end{center}
\caption{The supercooling scale $\Delta = 1 -
T_\mathrm{f}/T_\mathrm{dec}$ depends on the surface tension $\sigma$
for various models, which is estimated by
Eq.~(\ref{eqn:supercooling_scale}). The notations are as in
Fig.~\ref{fig:nucleationRatePrefactor_vs_supercooling}, except the
shadow regions around the lines show the difference between
Eq.~(\ref{eqn:supercooling_scale}) and the rough criterion $\Gamma
\simeq 1/d_\mathrm{H}^4$. The two gray vertical dashed lines are
marked for $\sigma = 0.02~T_\mathrm{dec}^3$ and $\sigma =
0.2~T_\mathrm{dec}^3$, which are chosen as typical parameters in
Fig.~\ref{fig:dnuc_vs_latentHeat_GubserModel}
and~\ref{fig:supercooling_vs_latentHeat_GubserModel}. It can be seen
that the supercooling scale $\Delta$ is really unsensitive to the
method we estimate it, even in the small $\sigma$ regions where
$d_\mathrm{nuc} \ll d_\mathrm{H}$. The thick lines are models (7
$\rightarrow$ 2 $\rightarrow$ 3 $\rightarrow$ 5 $\rightarrow$ 4
$\rightarrow$ 1) respectively seeing from the arrow direction.
Although $v_\mathrm{sh}$ can be calculated accurately
by~\cite{Kajantie:1992uk}, we choose $v_\mathrm{sh} = c_\mathrm{s}$
for simplification, where the differences between them are
imperceptible.} \label{fig:supercooling_vs_sigma}
\end{figure}

\begin{figure}[ht]
\begin{center}
\includegraphics[angle=0,width=9cm]{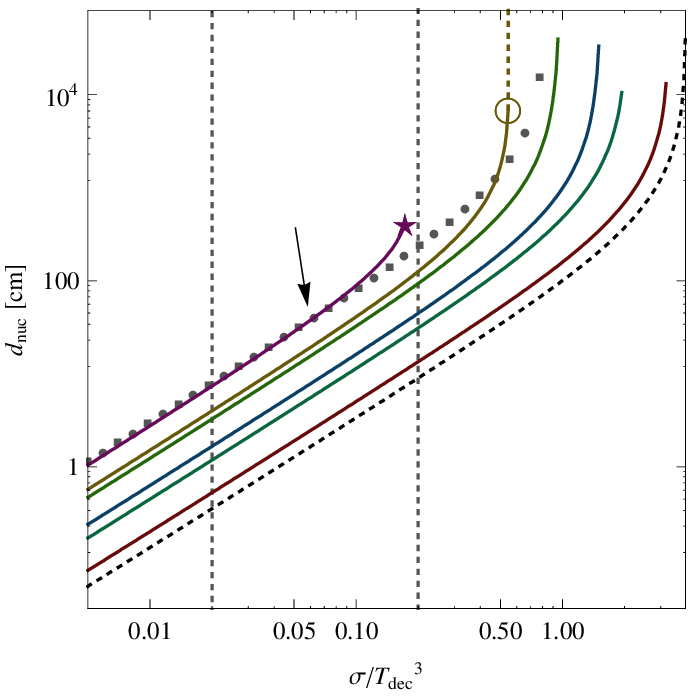}
\end{center}
\caption{The mean nucleation distance $d_\mathrm{nuc}$ depends on
various surface tension $\sigma$, estimated by Eq.~(\ref{eqn:dnuc}).
The notations are as in
Fig.~\ref{fig:nucleationRatePrefactor_vs_supercooling}. The thick
lines are models (7 $\rightarrow$ 2 $\rightarrow$ 3 $\rightarrow$ 5
$\rightarrow$ 4 $\rightarrow$ 1) respectively seeing from the arrow
direction. Although the terminal point ``$\star$'' marked for model
(7) is factual, the terminal point ``$\circ$'' marked for model (2)
is the numerical limit of our calculation. A maximum $\sigma$ exists
for the maximum expected supercooling scale to be achieved; as $L_h
= 0$ for $\Delta = \Delta_<$ in model (2), $d_\mathrm{nuc}
\rightarrow \infty$ when $\sigma$ tending towards this limit.}
\label{fig:dnuc_vs_sigma}
\end{figure}

The numerical result of $\Delta = 1 - T_\mathrm{f}/T_\mathrm{dec}$
depends on various surface tension $\sigma$ for various models, is
shown in Fig.~\ref{fig:supercooling_vs_sigma}. For small $\sigma$,
the system follows nicely to the relation $\Delta \propto
\sigma^{3/2}/L_\mathrm{h}$~\cite{Fuller:1987ue} for fixed
$L_\mathrm{h}$; however, when $\sigma$ is large enough, these lines
tilt up. One reason for these departures from $\sigma^{3/2}$ can be
seen from the reduction of Eq.~(\ref{eqn:supercooling_scale}) for
some EoS's with constant $L_\mathrm{h}$, which gives $\Delta \propto
\sigma^{3/2} / \sqrt{171 - 4 \ln{(\beta/\sigma^{3/2})}}$ for some
explicitly written positive $\beta$~\cite{Kajantie:1992uk}. The
other reason is the effective latent heat $L_\mathrm{h}$ released
drops for some not-very-small supercooling scale for the more
realistic EoS's. Nevertheless, comparing to the tilting up of
$d_\mathrm{nuc}$ seen form Fig.~\ref{fig:dnuc_vs_sigma}, the effects
here for $\Delta$ is really weak. Lines in that figure cannot be
extended to larger $\Delta$, in where $d(\Delta F/T)/d T \rightarrow
0$ and our approximation becomes inapplicable. In addition, $\Delta$
is totally insensitive to the prefactor in the right hand side of
Eq.~(\ref{eqn:supercooling_scale}), such as the shear viscosity
$\eta_\mathrm{q}$ or the shock viscosity $v_\mathrm{sh}$.

In a more accurate (and also more sophisticated) way, supercooling
scale can be calculated dynamically from the time evolution of the
temperature~\cite{Kapusta:2000fe}. We don't calculate the
time-dependent solutions here, because our qualitative QCD theories
still have too many free parameters, thus intrinsic discussions are
not very easy. Nevertheless, we think that there should be some
interesting results in the not-very-small supercooling regions.

The mean nucleation distance of the hadronic bubbles in the phase
transition era, can be estimated by $d_\mathrm{nuc} \simeq
n(t_\mathrm{f})^{-1/3}$ and the bubble number density calculated
in~\cite{Guth:1979bh,Guth:1981uk}. Some suitable reductions give
$d_\mathrm{nuc} \simeq (8\pi)^{1/3} v_\mathrm{sh} / (-d(\Delta
F/T)/d t|_\mathrm{t_\mathrm{f}})$~\cite{Ignatius:1994fr}.
Considering some special EoS, we have
\begin{equation}
    d_\mathrm{nuc} \simeq
        \left. \frac{(8\pi)^{1/3} v_\mathrm{sh}}{\sqrt{24 \pi G}}
        \frac{d \epsilon_\mathrm{q} / d T}{\epsilon_\mathrm{q}^{1/2} (p_\mathrm{q} + \epsilon_\mathrm{q})}
        \frac{d T}{d(\Delta F/T)} \right|_{T_\mathrm{f}} \mbox{.}
        \label{eqn:dnuc}
\end{equation}
The numerical result of $d_\mathrm{nuc}$ is shown in
Fig.~\ref{fig:dnuc_vs_sigma}. It can be seen that for small
$\sigma$, $d_\mathrm{nuc} \propto \sigma^{3/2}/L_\mathrm{h}$ for
fixed $L_\mathrm{h}$, as is estimated
in~\cite{Fuller:1987ue,Christiansen:1995ic}; however, when $\sigma$
becomes large, $d_\mathrm{nuc}$ tilts up caused by both a more
accurate treatment of supercooling and the drop of $L_\mathrm{h}$
for some more realistic EoS's. Although models (2) and (7) both have
some maximum $\sigma$ where $\Delta_<$ is saturated, their behavior
are completely different. In model~(2), $L_\mathrm{h} \rightarrow 0$
hence $d_\mathrm{nuc} \rightarrow \infty$ while $\Delta \rightarrow
\Delta_<$; but in model~(7), $L_\mathrm{h} \neq 0$ hence
$d_\mathrm{nuc}$ is finite.

\subsection{The Supercooling Scale and the Mean Nucleation Distance Once More}

It may not be plausible to consider the dependence of the
supercooling scale $\Delta$ and the mean nucleation distance
$d_\mathrm{nuc}$ on the surface tension $\sigma$. New phenomena
deviating from the rough analytic estimations $\Delta \propto
\sigma^{3/2}/L_\mathrm{h}$ and $d_\mathrm{nuc} \propto
\sigma^{3/2}/L_\mathrm{h}$~\cite{Fuller:1987ue}, always appear in
the regions where $\sigma$ is large enough. Although $\sigma$ is
indeed a free parameter since we do not know its value, it should
not be very large both from the lattice results of
gluodynamics~\cite{Iwasaki:1993qu,Beinlich:1996xg,Lucini:2005vg} and
some theoretical estimations based on the MIT bag
model~\cite{Farhi:1984qu,Berger:1986ps}. This issue has already been
discussed in Sec.~\ref{subsec:surface_tension}.

Notwithstanding, we can still do some qualitative or quantitative
estimations, and give some constraints on both the surface tension
$\sigma$ and the latent heat $L_\mathrm{h}$.

\subsubsection{The Global Constraint of the Surface Tension on the Large-$N_\mathrm{c}$ Theories
    \label{subsubsec:large-N_constraint}}

In~\cite{Lucini:2005vg}, the authors argued one cannot distinguish
the scaling of the surface tension $\sigma \propto N_\mathrm{c}$ or
$\sigma \propto N_\mathrm{c}^2$ from their lattice analyses of
$SU(N_\mathrm{c})$ gauge theories. However, for the reason that we
definitely know the latent heat $L_\mathrm{h} \propto
N_\mathrm{c}^2$ for a first order phase transition, if this
transition indeed exists, to avoid a zero nucleation rate in
Eq.~(\ref{eqn:nucleation_rate}), we need at most $\sigma \propto
N_\mathrm{c}^{4/3}$.

If in some large-$N_\mathrm{c}$ theories, $\sigma$ dependents on
$N_\mathrm{c}$ sharper than $N_\mathrm{c}^{4/3}$, we can
equivalently give an upper limit for $N_\mathrm{c}$. For the finite
temperature Scherk-Schwarz compactification model, the domain wall
tension $\sigma \propto \epsilon_\mathrm{q}/T_\mathrm{dec} \propto
N_\mathrm{c}^2$ has been calculated
numerically~\cite{Aharony:2005bm} for the compactified $AdS_5$ and
$AdS_6$ soliton solutions. Hence, given an explicit expanding
universe, we can restrict $N_\mathrm{c}$ by the phase transition
happened there. A special example to constrain $N_\mathrm{c}$ of the
large-$N_\mathrm{c}$ CFT in the holographic Randall-Sundrum (RS) I
model, is given in~\cite{Creminelli:2001th,Kaplan:2006yi}, despite
of the fact that the concept of the surface tension does not
intervene their discussions. The exponential suppressive factor in
the nucleation rate formula, is given by the Euclidean action which
has a minimum at $T = 1/\sqrt{3}~T_\mathrm{c}$ for some transition
happens at $T_\mathrm{c}$. The comparison between the holographic RS
I phase transition and our model based on AdS/CFT, is given in
Sec.~\ref{sec:conclusion}.

\subsubsection{The Extremely Weakly First Order Confinement/Deconfinement Phase transition?
    \label{subsubsec:extremely_weak_1st_order}}

The order of the confinement/deconfinement phase transition for QCD
with physical quark masses, is still being debated. The lattice
results of quenched QCD suggest that it is at most weakly first
order~\cite{Beinlich:1996xg}. However, adding massive quarks seems
to make the transition weaker, or even gradually changing it to a
rapid crossover~\cite{Fodor:2004nz,Bernard:2004je}. Hence, one
possibility to be considered is the extremely weakly first order
case. We still assume the bubbles expand deflagratingly in this
case.

Na\"{i}vely, both the supercooling scale $\Delta$ and the mean
nucleation distance $d_\mathrm{nuc}$ increase reciprocally while the
latent heat $L_\mathrm{h}$ decreases, base on the rough analytic
estimations $\Delta \propto \sigma^{3/2}/L_\mathrm{h}$ and
$d_\mathrm{nuc} \propto
\sigma^{3/2}/L_\mathrm{h}$~\cite{Fuller:1987ue}. However, more
abundant phenomena can happen for more realistic EoS's of QCD.

\begin{figure}[ht]
\begin{center}
\includegraphics[angle=0,width=12cm]{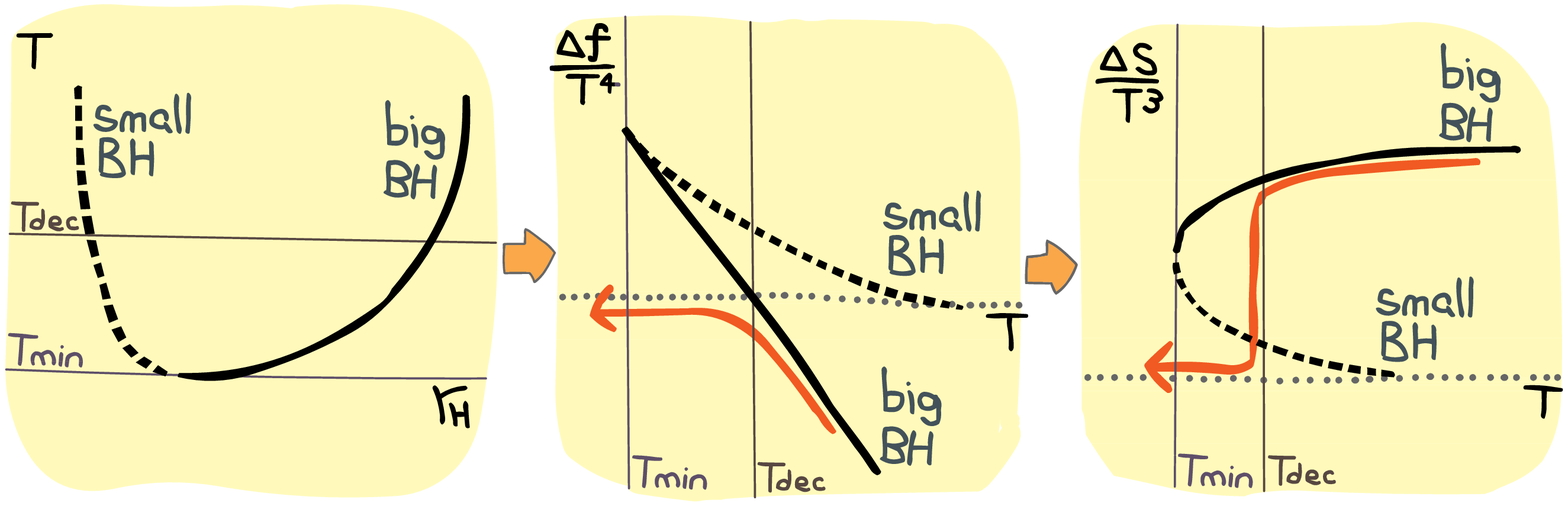}
\end{center}
\caption{A Hawking-Page phase transition~\cite{Hawking:1982dh}
should always have a minimum temperature $T_\mathrm{min}$, below
which the high temperature phase cannot exist. This minimum
temperature is intrinsic, rather than caused by impurities or
perturbations in the old supercooling scenarios. The long curved
arrows show the behavior of the system from high temperature to low
temperature phase, if no supercooling happens.}
\label{pic:Hawking-Page_minimumT}
\end{figure}

These phenomena are caused mainly by two reasons. (i) If the EoS's
possesses the weakly first order phase transitions, the effective
$L_\mathrm{h}$ decreases when the supercooling scale becomes large.
This can easily be seen from Fig.~\ref{fig:T_vs_S} and the relation
$L_\mathrm{h} = (4/3) T (s_\mathrm{q} - s_\mathrm{h})$. (ii) As a
universal property of the Hawking-Page phase
transition~\cite{Hawking:1982dh}, there is a minimum temperature
$T_\mathrm{min} < T_\mathrm{dec}$ below which the high temperature
phase cannot exist. It is illustrated in
Fig.~\ref{pic:Hawking-Page_minimumT}. The qualitative effect of the
first reason has already been discussed in~\cite{Ignatius:1994fr}.
We will give here both quantitative effects of (i) for some specific
EoS's, and also some qualitative effects of (ii).

For our discussions, we will use the mimicking model of Gubser
\emph{et al.}~\cite{Gubser:2008ny,Gubser:2008yx,Gubser:2008sz}. The
reason is that, it is convenient to use its potential $V(\phi)$ to
construct a first order phase transition with decreasing
$L_\mathrm{h}$, which then transforms smoothly to a rapid crossover.
Another phenomenological model including a dilaton field given
in~\cite{Gursoy:2008bu,Gursoy:2008za} may also be used, as it has a
more solid theoretical foundation. We omit the discussions of it
here, because the work for this model itself is still on its way,
and the calculation of the EoS's is more complicated than the former
one. Some qualitative properties, such as $\Delta_<$ decreases with
decreasing $L_\mathrm{h}$, are supposed to be universal.

\begin{figure}[ht]
\begin{center}
\includegraphics[angle=0,width=9cm]{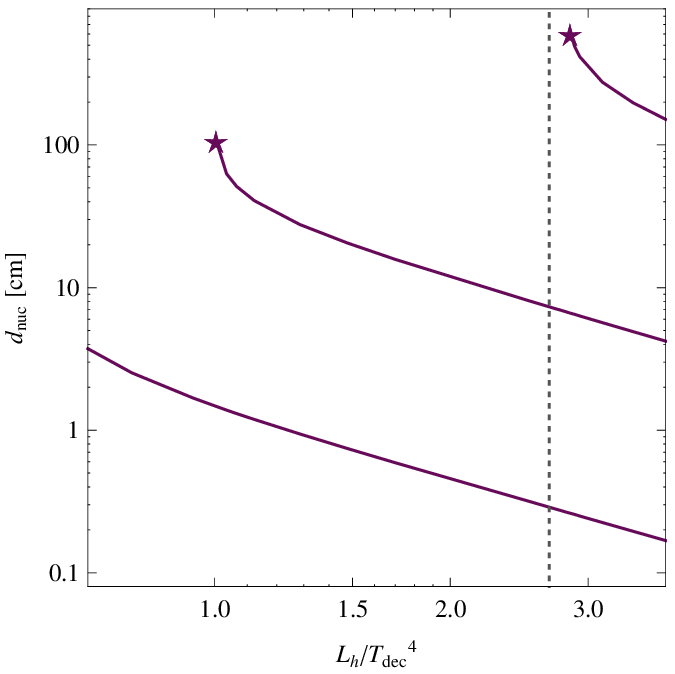}
\end{center}
\caption{The mean nucleation distance $d_\mathrm{nuc}$ depends on
the latent heat $L_\mathrm{h}$ for the Gubser \emph{et al.}
model~\cite{Gubser:2008ny}. The gray vertical dashed line is
$L_\mathrm{h} = 2.67~T_\mathrm{dec}^4$ deduced from the potential
$V(\phi) = [-12 \cosh{(\sqrt{7/12} \phi)} + 2 \phi^2] / L^2$ in the
formal estimations. The three thick lines are for $\sigma =
0.2~T_\mathrm{dec}^3$, $0.02~T_\mathrm{dec}^3$ and
$0.002~T_\mathrm{dec}^3$ (from up down), respectively.}
\label{fig:dnuc_vs_latentHeat_GubserModel}
\end{figure}

\begin{figure}[ht]
\begin{center}
\includegraphics[angle=0,width=9cm]{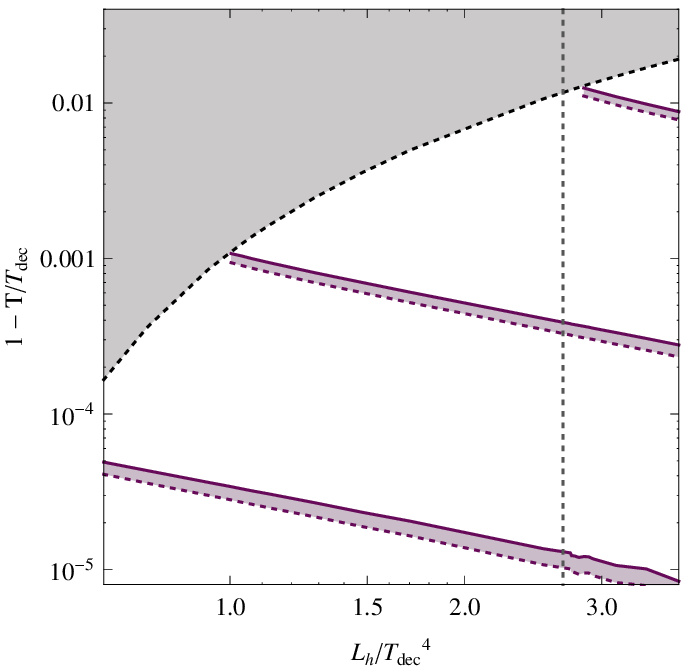}
\end{center}
\caption{The various supercooling scales depend on the latent heat
$L_\mathrm{h}$ for the Gubser \emph{et al.}
model~\cite{Gubser:2008ny}. The black dotted curve from the
top-right corner to the bottom-left corner is the maximum
supercooling scale $\Delta_<$; hence the shadow region above it, is
forbidden by that model. The actual supercooling scales $\Delta = 1
- T_\mathrm{f}/T_\mathrm{dec}$ calculated by Eq.~(\ref{eqn:dnuc})
are denoted by the thick solid lines, which are for $\sigma =
0.2~T_\mathrm{dec}^3$, $0.02~T_\mathrm{dec}^3$ and
$0.002~T_\mathrm{dec}^3$ (from up down) respectively. The dotted
lines a little below them constrain the phase transition to be
completed, which are roughly calculated by $\Gamma \simeq
1/d_\mathrm{H}^4$; that is, $d_\mathrm{nuc} \simeq d_\mathrm{H}$.
For some particular $\sigma$, $d_\mathrm{nuc}$ can easily be much
larger, providing that the latent heat $L_\mathrm{h}$ is small
enough that the maximum supercooling $\Delta_<$ is saturated.
However, it is unlikely that the larger $d_\mathrm{nuc}$ can help us
understanding the formation of quark nuggets or the inhomogeneous
initial conditions of the big-bang nucleosynthesis, because the
parameter $L_\mathrm{h}$ should be fine tuned.}
\label{fig:supercooling_vs_latentHeat_GubserModel}
\end{figure}

As the potential of the Gubser \emph{et al.} model $V(\phi) = [-12
\cosh{(\gamma \phi)} + b \phi^2] / L^2$ has two parameters $\gamma$
and $b$, in fact, our method applies to a wide range of models
(potentials) with one free parameter. We fix $b = 2$ and evaluate
$\gamma \in [0.722,~0.790]$ (formerly we used $\gamma = \sqrt{7/12}
\simeq 0.764$); when doing this, the latent heat $L_h$ varies from
$0.69$ to $3.77~T_\mathrm{dec}^4$. The dependence of
$d_\mathrm{nuc}$ and the supercooling scale $\Delta$ on
$L_\mathrm{h}$ are shown in
Fig.~\ref{fig:dnuc_vs_latentHeat_GubserModel}
and~\ref{fig:supercooling_vs_latentHeat_GubserModel}. Observing from
the figures, when $L_\mathrm{h}$ is large enough, it follows the
scaling law $\Delta \propto L_\mathrm{h}^{-1}$ and $d_\mathrm{nuc}
\propto L_\mathrm{h}^{-1}$; however, for smaller $L_\mathrm{h}$,
$d_\mathrm{nuc}$ tilts up because the effective $L_\mathrm{h}$ drops
for reason (i). In a large acceptable parameter space,
$d_\mathrm{nuc}$ is not as small as people used to think as about
$\simeq 2~\mathrm{cm}$~\cite{Christiansen:1995ic} for the
homogeneous nucleation case. For some definite $\sigma$, there exist
some minimum $L_\mathrm{h,<}$, where the maximum supercooling
$\Delta_<$ is achieved.

What happens if the realistic $L_\mathrm{h}$ is smaller than
$L_\mathrm{h,<}(\sigma)$? Maybe this situation never happens in a
consistent world. In despite of that, as a lack of the complete
origin of the surface tension, we just treat $L_\mathrm{h}$ and
$\sigma$ as free parameters. If this happens, we have the bubble
number density
\begin{equation}
    n(t_\mathrm{<}) \simeq
        \left. \left( \frac{\kappa}{2\pi} \Omega_0 \right) \frac{e^{-\Delta F/T}}{-d(\Delta F/T)/dt}\right|_{t=t_<}
        \ll \left.\frac{[-d(\Delta F/T)/dt]^3}{8 \pi v_\mathrm{sh}^3}\right|_{t=t_<}
        \mbox{,}
\end{equation}
comparing with Eq.(\ref{eqn:dnuc}) and the discussions
in~\cite{Ignatius:1994fr}, where $t_<$ is the time when the minimum
temperature $(1-\Delta_<)T_\mathrm{dec}$ is achieved. Because of the
exponential suppressed factor $\exp{(-\Delta F/T)}$, this situation
will lead to a much smaller bubble number density $n$ hence a much
larger $d_\mathrm{nuc}$. One may think that the larger
$d_\mathrm{nuc}$ can help surviving the quark nuggets, or provide
the inhomogeneous initial conditions of the big-bang nucleosynthesis
(BBN). However, this scenario is in fact rather hard to appear. We
also show in Fig.~\ref{fig:supercooling_vs_latentHeat_GubserModel}
the criterion $\Gamma \simeq 1/d_\mathrm{H}^4$, that is, the
supercooling scale needed for $d_\mathrm{nuc} \simeq d_\mathrm{H}$.
Because $d_\mathrm{nuc}$ varies too sensitively to the supercooling
scale, the corresponding $L_\mathrm{h}$ has some value very close to
$L_\mathrm{h,<}$. Hence, to get an appropriate $d_\mathrm{nuc}$ for
our universe, we need to fine-tune $L_\mathrm{h}$ in a very small
region a little smaller than $L_\mathrm{h,<}$, which is unlikely to
be so.

\section{Discussion and Conclusion\label{sec:conclusion}}

In this paper, we discussed some implication of the new AdS/CFT
results to the cosmological QCD confinement/deconfinement phase
transition. We limit our discussion to the homogeneous nucleation
case. The values of the hydrodynamical quantities, like the shear
viscosity $\eta$ or the bulk viscosity $\zeta$, can significantly
lower the prefactor $(\kappa/2\pi) \Omega_0$ of the nucleation rate
formula compared to the old estimations; however, they can hardly
affect other characteristic parameters of this process, such as the
supercooling scale $\Delta = 1 - T_\mathrm{f}/T_\mathrm{dec}$ or the
main nucleation distance $d_\mathrm{nuc}$. The new EoS's, which
differ from the MIT bag model, can affect the phase transition
scenario mainly in two ways. (i) As most of these EoS's are
comparatively more weakly first order than the bag model, it is not
adequate to treat their latent heat $L_\mathrm{h}$ as a constant.
For some not-very-small supercooling, the effective latent heat is
always much smaller. Hence, $d_\mathrm{nuc}$ enhances comparing to
the old estimation $d_\mathrm{nuc} \propto
\sigma^{3/2}/L_\mathrm{h}$~\cite{Fuller:1987ue} when $\sigma$
becomes larger or $L_\mathrm{h}$ becomes smaller. In a large
acceptable parameter space of $\sigma$ and $L_\mathrm{h}$,
$d_\mathrm{nuc}$ is not as small as people used to think as about
$\simeq 2~\mathrm{cm}$~\cite{Christiansen:1995ic} for the
homogeneous nucleation case. (ii) The high temperature phase should
have an intrinsic maximum supercooling scale $\Delta_<$ based on a
Hawking-Page type phase transition. This is in contrast with the old
belief that the range of supercooling is caused by impurities or
perturbations. We discussed the possibility that this maximum
supercooling scale is saturated in the cosmological QCD phase
transition, which may happen when this phase transition is extremely
weakly first order. If it happens, the nucleation distance
$d_\mathrm{nuc}$ can be increased tremendously. However, it is
unlikely to be so; because to get an appropriate $d_\mathrm{nuc}$
for our universe (that is, to help understand the surviving of the
quark nuggets, or to get the appropriate initial conditions of the
BBN), $L_\mathrm{h}$ needs to be fine tuned.

Some related works are listed as below for comparison reasons. The
nucleation rate and also some of its cosmological applications, base
on the holographic RS I model, are discussed
in~\cite{Creminelli:2001th,Kaplan:2006yi}. In this model, a ``Planck
brane'' and a ``TeV brane'' are added to the $AdS_5 \times S^5$
spacetime with a dual CFT. The ``Planck brane'' makes a UV cutoff
hence adds a $(3+1)$-dimensional gravity; the ``TeV brane'' makes an
IR cutoff, and the standard model fields in it are understood as
bound states out of the strong interacting
CFT~\cite{ArkaniHamed:2000ds}. When at finite temperature, to make a
lower free energy, the low temperature phase is as in the RS I
model, but the high temperature phase favors an AdS-Schwarzschild
solution (duals to the free CFT gas); hence, our universe should
suffer a phase transition at some $T_\mathrm{c}$ lower than the
Fermi scale. To ensure that the phase transition is completed thus
for avoiding an empty universe, we need a strong upper bound for
$N_\mathrm{c}$ of the dual CFT field. This model has already been
discussed in Sec.~\ref{subsubsec:large-N_constraint}, where we
pointed out that an upper limit of $N_\mathrm{c}$ may be universal
for some large-$N_\mathrm{c}$ theories which suffer some phase
transitions.

The phase transition of an AdS/CFT model, in which a
$(2+1)$-dimensional field theory is dual to some (confined) AdS
soliton or some (deconfined) black 3-brane metric compactified in a
brane dimension, is discussed in~\cite{Horowitz:2007fe}. The
supercooling and the rapid reheating (hadronization) after it, are
considered. Notwithstanding, in the large-$N_\mathrm{c}$ limit, the
slowly hadronized phase at the temperature $T_\mathrm{dec}$ do not
happen in their model. To begin at some supercooling temperature
$T_0 > 0$, the residual deconfined regions after the phase
transition still hold the energy portion larger than $1/4$. In that
model, the supercooling scale is given by hand, and a lower limit
$T_0 = 0$ ($\Delta = 1$) is considered. Comparing to that work, what
we do in this paper is calculating $\Delta$ explicitly within some
physical environments (what we use is the cosmological QCD phase
transition). We use some AdS/CFT models more pertinent to the
$(3+1)$-dimensional QCD than theirs.

In addition, an interesting relation between the KSS bound and
strange quark stars, is shown in~\cite{Bagchi:2007ir}. The authors
argued that, the surface of quark stars at the temperature $T \sim
80~\mathrm{MeV}$, has already saturated the KSS bound.

The question which parallels to the topic we discussed in this
paper, is how the RHIC results of strong interacting QGP and the
AdS/CFT correspondence can affect the research of neutron stars and
quark stars. The difference is that the deconfined QGP in quark
stars is mainly caused by its high chemical potential, rather than
caused by their high temperature in RHIC or the early universe. A
lot of AdS/CFT models for finite chemical potential have already
been constructed; although just as in the finite temperature case,
they are mainly studied in the large-$N_\mathrm{c}$ limit. We will
leave these issues to the follow-up studies.

\section*{Acknowledgements}

I would like to thank Ofer Aharony, Oleg Andreev, Thomas Cohen,
Joshua Erlich, Itzhak Fouxon, Umut G\"{u}rsoy, Christopher Herzog,
Keijo Kajantie, Joseph Kapusta, David Mateos, Berndt M\"{u}ller,
Robert Myers, Francesco Nitti, Matthew Roberts, Shigeki Sugimoto and
Xin-Nian Wang for helpful discussions of issues related to this
paper, and Gang Chen for organizing the folk AdS/CFT seminars.
Darren Shih read the manuscript and gave me some advice and
suggestions.

\bibliographystyle{bib_style}
\bibliography{../0810_AdS-CFT,../0810_phase_transition,../0810_QGP}


\end{document}